\documentclass{easychair}

\usepackage{url}
\usepackage{doc}
\usepackage[printonlyused,withpage]{acronym}
\title{BAMSim Simulator}

\author{
Rafael F. Reale\inst{2}
\and Walter P. Neto\inst{1}
\and Joberto S. B. Martins\inst{1}
}

\institute{
  Salvador University (UNIFACS),
  \email{wcpneto@gmail.com, joberto.martins@unifacs.br}
\and
   Federal Institute of Bahia (IFBA),
   \email{reale@ifba.edu.br}
}

\authorrunning{Reale and Martins}

\titlerunning{ADVANCE 2021}

\begin{document}

\maketitle

\begin{abstract}
Resource allocation is an essential design aspect for current  systems and bandwidth allocation is an essential design aspect in multi-protocol label switched and OpenFlow/SDN network infrastructures. The bandwidth allocation models (BAMs) are an alternative to allocate and share bandwidth among network users. BAMs have an extensive number of parameters that need to be defined and tuned to achieve an expected network performance. This paper presents the BAMSim simulator to support the network manager decision process in choosing a set of BAM configuration parameters for network design or during network operation.

\end{abstract}

\section{Introduction}
\label{sect:introduction}

The allocation of resources is an essential aspect in designing existing systems like the internet of things (IoT), smart cities, smart grid, and new generation 5G and 6G systems \cite{liu_computation_2020}. Bandwidth is a resource allocated for network substrates that use links between switching nodes to support user communications. Bandwidth allocation is an essential aspect of design in broadly used network infrastructures like multi-protocol label switching (MPLS) network and network infrastructure deployment approaches using OpenFlow/SDN to control level-2 switches or any other switching equipment \cite{li_enhancing_2020}.

The bandwidth allocation models (BAMs) are an alternative to allocate and share bandwidth among network users grouped in traffic classes representing their communication requirements \cite{martins_uma_2015}. BAMs are mostly used in networks with limited bandwidth in which there is no over-provisioning for the communication links \cite{reale_analysis_2014}.

This paper presents the BAMSim bandwidth allocation model (BAM) simulator to support the network manager decision process in choosing an appropriate set of BAM configuration parameters during network design and operation.

\section{BAMSim Simulator}\label{sec:BAMSim}

The BAMSim is specifically oriented to support the allocation of bandwidth on MPLS networks according to the DS-TE (DiffServ Aware - Traffic Engineering) style proposed by Faucher et al. \cite{faucher_maximum_2005}. In summary, the DS-TE style means that bandwidth is allocated by traffic classes (TC), with each TC having a bandwidth constraint (BC).

The BAMSim processes every new LSP demand for the network. The BAM model decides if the LSP is created (allocating bandwidth) or rejected. LSP's requests can be granted, blocked, or preempted depending on the traffic class bandwidth available. Abstractly, an LSP has its starting time, duration, TC, and required bandwidth. The execution module is optionally available to interface the BAMSim simulator with a physical or an emulated network like the MiniNet with OpenFlow-based network control.

BAMSim facilities include: i) The support of MPLS signaling protocol features like LSP establishment, LSP teardown, LSP block and LSP preemption; ii) Network topology definition; iii) Path computation with static matrix and Constrained Shortest Path First (CSPF) route computation; and iv) Input traffic generation (with poisson, exponential and uniform distribution and deterministic traffic).

BAMSim supports the basic BAM models Maximum Allocation Model (MAM) \cite{faucher_maximum_2005}, Russian Dolls Model (RDM) \cite{le_faucheur_russian_2005}, and AllocTC-Sharing (ATCS) \cite{reale_alloctc-sharing:_2011} and the generalized BAM model (GBAM) \cite{reale_g-bam:_2014}. The advantages of BAM dynamic model switching and reconfiguration are discussed in \cite{reale_analysis_2014}  \cite{reale_applying_2016}.

The BAMSim has a modular internal structure implemented in Java. The BAMSim simulator is available at \url{https://github.com/rfreale/BAMSim}.

BAMSim related work includes the simulation of BAM models based on the NS (Network Simulator) presented in  Adami et al.  \cite{adami_new_2008}. The NS module developed simulates MAM and RDM BAM models. Compared with Adami's NS-based module, the BAMSim simulator extends the simulation of BAM modules to all existing models. To the limit of our knowledge, no other BAM simulator as developed since then.

 \section{BAMSim Architecture and Operation} \label{sec:arch}

\begin{figure}[tb]
	\begin{centering}
	\includegraphics[width=0.7\textwidth]{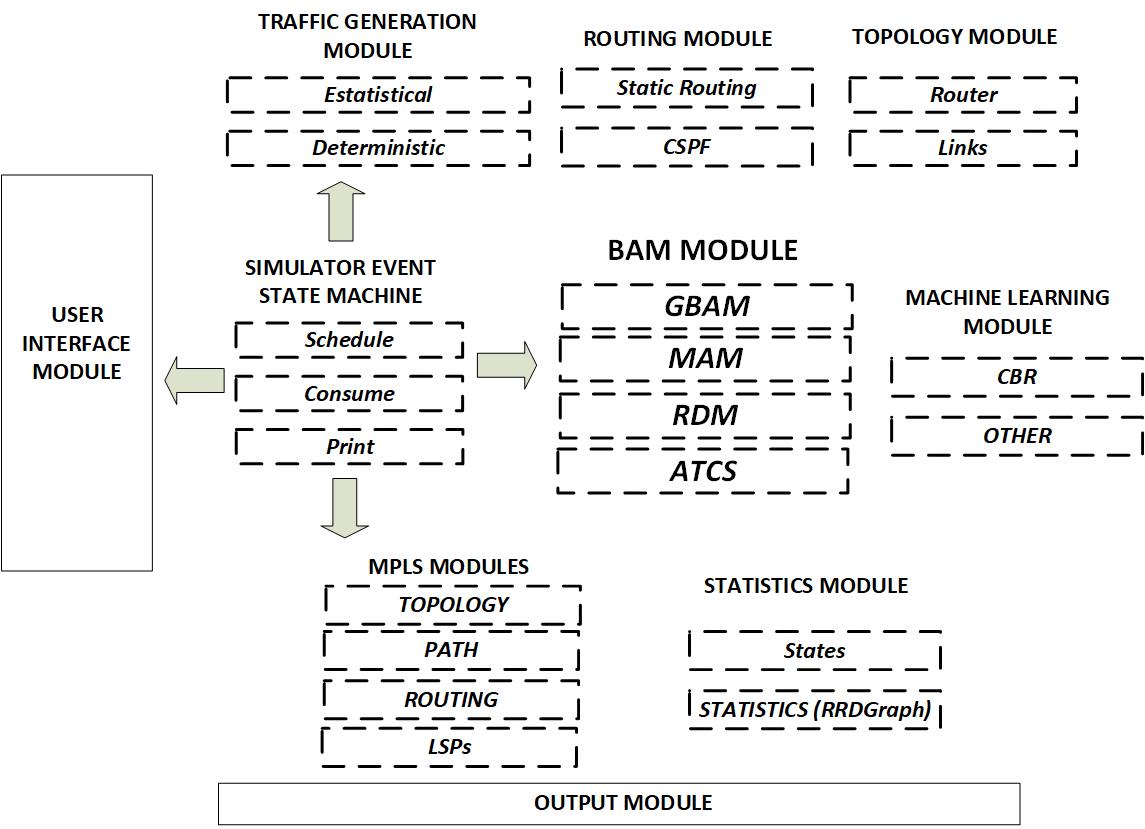}
	\caption{BAMSim Architecture and Internal Modules}
	\label{fig:BAMSIM_Modules}
	\end{centering}
\end{figure}

The BAMSim architecture and main modules are illustrated in Figure \ref{fig:BAMSIM_Modules}.

BAMSIM architecture is modular, configurable, and flexible. The simulator code allows the inclusion of functionalities by integrating new modules. The BAMSim allows the configuration of various simulation parameters like simulation time, the simulation runs, number of generated LSPs, LSP duration time, simulation stop conditions, and pseudo-aleatory seeds for traffic generation. BAMSim flexibility includes its capability to import topologies like NSFNet, NTT (Nippon Telegraph and Telephone Corporation), define file-based customized topologies and define routing in the MPLS network with static matrix or protocol computed routing based in protocols like the CSPF.

The MPLS modules abstractly implement the main functions of an MPLS network such as establishment, preemption, devolution, and blocking of LSPs by traffic class (TC) through its interaction with the BAM, topology, path and routing modules in a configurable way.

The BAM module implements the BAM models. A GBAM-based implementation was used in the BAMSim to allow the implementation of all existing BAM models with its operational characteristics and behavior. Another important aspect of using GBAM is that it allows the on-the-fly switching of BAM models to adequate input traffic dynamism, as discussed in \cite{reale_analysis_2014}.

The cognitive module supports the use of machine learning techniques in assisting BAM parameters configuration and BAM model switching according to network traffic and state conditions and according to the objectives defined by the network manager. In the current BAMSim implementation, a Case-based Reasoning (CBR) module is implemented using the jColibri framework \cite{oliveira_cognitive_2018}.

The topology module represents the MPLS network routers and links in an abstract way. The module represents routers interconnection as well as link bandwidth and active traffic classes per link. Topologies can be configured manually or imported by text files.

The routing module enables path selection by integrating a path selection algorithm. The BAMSim can use the CSPF or manually configured path selection algorithms. The routing module is developed in rJava, which allows us to explore the R language resources.

The status and statistics module monitors the states of the network elements and generates statistics and graphics with a Round Robin Database (RRD) using the jRobin library. Statistics are recorded in RRD files since RRD data and charts are intuitive for network managers and used in traditional network monitoring tools.

The traffic generator module allows BAMSim to generate random traffic profiles through probability functions such as uniform, Poisson, exponential, logarithmic, and deterministic traffic for DEBUG and traffic imported from real networks.

\section{BAMSim Simulator Application Scenarios}\label{sec:scenarios}

Using BAMSim for BAM-based network design tuning means to evaluate how the configured input traffic impacts the MPLS network performance parameters like LSP preemption, LSP blocking, and link utilization for distinct BAM models. Table \ref{tab:TrafficProfile} illustrates a simulation scenario.

\begin{table}[htbp]
\small
\centering
\caption{Input Traffic - Simulation Pattern Example}
%\vspace*{+5mm}
\label{tab:TrafficProfile}
\begin{tabular}{|c|c|c|c|c|c|c|}
\hline
%\small
 \textbf{Traffic Profile} & \textbf{1}	& \textbf{2} & \textbf{3} & \textbf{4}	& \textbf{5} & \textbf{6}  \\ \hline
TC0 & High & Medium & Low & \multicolumn{3}{|c|}{High} \\\hline
TC1 & Low & Low & Medium & \multicolumn{3}{|c|}{High}\\\hline
TC2 & Low & High & High & \multicolumn{3}{|c|}{High}\\ \hline
Link utilization & \multicolumn{3}{|c|}{$<$ 90\%} & \multicolumn{3}{|c|}{$>=$ 90\%} \\ \hline
%BAM Indicado & RDM/ATCS & ATCS & ATCS & \multicolumn{3}{|c|}{MAM} \\ \hline
\end{tabular}
\end{table}

The simulation defines six input traffic profiles with 1 hour each. The first three traffic profile combinations alternate the need for sharing bandwidth between TCs and allows to assess what is the best BAM model option (MAM, RDM or ATCS) for each traffic profile. The last three input traffic profiles force link overload and allow to verify network performance parameters (blocking, preemption, other) and link utilization conditions for distinct BAM models and tune the one that best suits the network manager objectives.

Another BAM-based network-tuning possibility is to reconfigure the traffic class (TC) bandwidth constraints (BC) and, by simulation, to evaluate how the network performance parameters behave for different BAM models.

BAMSim configuration is a straightforward task executed by command line (text file scripts) and GUI-based instructions to the simulator. For example, consider the simulation for a point-to-point MPLS topology with two routers connected by a network link. BAMSim configuration steps include the definition of the network topology (file scripts), input traffic generation characteristics, routing matrix, simulation stop criteria, BAM model and BAM configuration parameters. Script command syntax and semantics are obviously specific to the BAMSim. For example, $PTP-2n-1e$ is the topology name with two nodes and one link between routers $0$ and $1$. Additional syntax and semantics information is available with the BAMSim code.

The BAMSim simulator's utilization to facilitate the learning curve of the BAM configuration process is another relevant application scenario. The BAM teaching simulator's focus is now to allow the modification of the BAM configuration parameters and visualize their effect on the network performance. An application program, based on the BAMSim, was developed with this purpose and is available at \url{https://github.com/rfreale/BAMSim/tree/Education}.

\section{Final Considerations} \label{sec:conclusion}

The BAMSim, in general, aims to facilitate the design process of bandwidth allocation in network infrastructures and foster the BAM-based bandwidth allocation learning curve. The BAMSim simulator facilitates the manager decision process in choosing the best BAM model and its operating parameters.

\label{sect:bib}
\bibliographystyle{plain}
\bibliography{allbib}

\end{document}